\documentclass[twocolumn,showpacs,prb]{revtex4}
\usepackage{amsmath}
\usepackage{amssymb}
\usepackage{graphicx}

\DeclareMathOperator{\arctanh}{artanh}

\renewcommand\Re{\mathop{\rm Re}}
\renewcommand\Im{\mathop{\rm Im}}

\begin{document}

\title{Effects of non-equilibrium noise on a quantum memory encoded in Majorana
zero modes}

\author{Fran\c{c}ois Konschelle and Fabian Hassler}

\affiliation{Institute for Quantum Information, RWTH Aachen University, 52056
Aachen, Germany}

\date{August 5, 2013}


\begin{abstract}
In order to increase the coherence time of topological quantum memories in
systems with Majorana zero modes, it has recently been proposed to encode the
logical qubit states in non-local Majorana operators which are immune to
localized excitations involving the unpaired Majorana modes.  In this
encoding, a logical error only happens when the quasi-particles, subsequent to
their excitation, travel a distance of the order of the spacing between the
Majorana modes.  Here, we study the decay time of a quantum memory encoded in
a clean topological nanowire interacting with an environment with a particular
emphasis on the propagation of the quasi-particles above the gap.  We show
that the non-local encoding does not provide a significantly longer coherence
time than the local encoding.  In particular, the characteristic speed of
propagation is of the order of the Fermi velocity of the nanowire and not
given by the much slower group velocity of quasi-particles which are excited
just above the gap.
\end{abstract}
\maketitle 

Since their introduction to condensed matter about a decade ago, Majorana zero
modes attract a lot of interests, especially regarding their quantum
information perspectives.\cite{Kitaev2001,sarma:05,Nayak2008}  On the one hand,
their non-Abelian statistics can be used to manipulate the quantum states,\cite{sarma:05,alicea:11,Sau2011a,VanHeck2012} opening interesting
possibilities in the recently proposed scheme of topological quantum
computation.\cite{Nayak2008}  On the other hand, the possibility to
efficiently store quantum information encoded in Majorana zero modes seems
very promising.\cite{Beenakker2011}

A Majorana zero mode is described by a self-adjoint operator
$\gamma_i=\gamma_i^\dag$.  Distinct Majorana modes obey the fermionic
anticommutation relations $\{ \gamma_i, \gamma_j\} = 2
\delta_{ij}$.\cite{Beenakker2011,alicea:12} Since
they break the U(1) symmetry of electric charge conservation down to
$\mathbb{Z}_2$, it is natural to search for them emerging in superconducting
systems where they appear as boundary states in chiral $p$-wave
nanowires.\cite{Kitaev2001} Even so there is no occupation operator associated
with a single Majorana mode due to the fact that
$\gamma{}^{\dagger}\gamma=\gamma^{2}=1$, two Majorana modes can be combined to
a single conventional fermionic mode $c= \tfrac12(\gamma_1 + i \gamma_2)$ with
the corresponding number operator $c{}^{\dagger}c =
\tfrac12(1+i\gamma_{1}\gamma_{2})$.  This fact in turn indicates that in
electronic systems emergent Majorana modes will always appear in pairs.
Surprisingly, a situation is possible where the two Majorana modes
$\gamma_{1}$ and $\gamma_{2}$ belonging to a single fermionic mode $c$ are
spatially separated from each other (unpaired), more precisely they are
totally delocalized at the two ends of a superconducting nanowire.  These two
delocalized modes $\gamma_{1}$ and $\gamma_{2}$ when taken together represent
a fermionic mode at zero energy which encodes the parity of the total number
of fermions in the system.  Because the fermion parity is a conserved quantity
for an isolated superconductor, a quantum state encoded in a wire hosting
Majorana modes is in principle immune to decoherence and thus serves as an
interesting implementation of a quantum memory.  On the other
hand, if the superconductor exchanges (quasi-)particles with its environment,
e.g., when in proximity to a gapless metal, the parity is not conserved and
there is no topological protection of the memory, see, e.g.,
Refs.~\onlinecite{leijnse,budich}.

\begin{figure} 
  \centering 
  \includegraphics[width=0.8\linewidth]{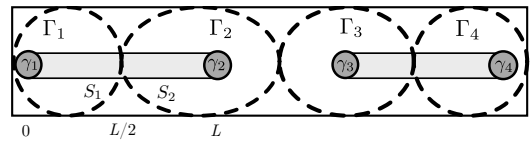}
\caption{The different options to encode a qubit into Majorana modes: the most
basic choice is to encode the qubit in the localized Majorana modes
$\gamma_{i}$, represented by gray disks at the end of the two light gray wires
which are located on top of a superconductor.  A local interaction pumping the
energy $\Delta$ into the system decoheres the qubit when the Majorana
wave-function excites a quasi-particle above the superconducting gap.  In the
macro-Majorana encoding, the Majorana modes are replaced by non-local
operators $\Gamma_{i}$ which involve the localized Majorana mode $\gamma_i$
and the parity of the number of fermions (quasi-particles) in the area $S_i$
surrounded by the dashed curves.  In this encoding the qubit only decoheres
when a Majorana mode excites a quasi-particle mode above the gap which
subsequently travels the distance $L/2$ (with $L$ the wire's length) where it
crosses into the next dashed region.  \label{fig:Q_memory_encoding}}
\end{figure}

Due to the superselection rule, superposition of different parity states are
unphysical.  Thus, in order to encode a qubit of information in the Majorana
modes of a topological wire, the previous picture has to be slightly modified:
in fact due to the conservation of the total fermion parity, four Majorana
modes $\gamma_{1}$, \dots, $\gamma_{4}$ at the edges of two wires are needed
to encode a single qubit (see Fig.~\ref{fig:Q_memory_encoding}).  As the total
fermion parity operator $P\equiv-\gamma_{1}\gamma_{2}\gamma_{3}\gamma_{4} = \pm 1$ is
a conserved quantity, the relative parity between the two wires encodes the
qubit state $Z=i\gamma_{1}\gamma_{2}= iP
\gamma_{3}\gamma_{4}$.\cite{Beenakker2011}

Since there is not any known example of a natural topological (chiral
$p$-wave) superconductor at the moment, Majorana modes have been proposed to
emerge in a closely related realization: a semiconducting nanowire with strong
spin-orbit effect, in a magnetic field, and proximity coupled to a
conventional ($s$-wave) superconductor.\cite{lutchyn:10,Oreg2010,referee2:10}  For this
geometry, the presence of Majorana modes may have already been observed last
year,\cite{mourik:12,deng:12,das:12,finck:13,churchill:13} see
Ref.~\onlinecite{Franz2013} for a discussion.

Nevertheless, the presence of a rather small proximity induced gap alters the
robustness of the quantum memory encoding: the zero-energy ground state is not
enough isolated to be efficiently protected, and the excitations of the
zero-energy modes above the energy gap destroy the quantum memory.
\cite{Goldstein2011,Schmidt2012} The failure of the encoding comes from the
absence of a topological protection for any local one-dimensional system at
non-vanishing temperatures.\cite{Dennis2002} If a perturbation is strong
enough to excite one of the localized zero-energy mode (say $\gamma_{1}$ for
instance) into an excited quasi-particle above the energy gap, the sign of the
corresponding Majorana mode flips resulting in a qubit sign-error.

To overcome this problem, Akhmerov recently proposed a non-local qubit
encoding, hereafter called a macro-Majorana encoding, which is in principle
robust to local excitations.\cite{Akhmerov2010} The robustness originates from
the localization of the excitations in a portion of space containing one of
the unpaired Majorana modes (see Fig.~\ref{fig:Q_memory_encoding} for a
schematic picture).  Then, the total system can formally be cut into distinct
sections $S_i$, each of them having only one Majorana mode $\gamma_i$.  As
long as the excitation quasi-particles do not enter into an adjacent region, a
 non-local Majorana operator $\Gamma_i = \gamma_i \prod_{x \in S_i}
(-1)^{c^\dag_x c_x}$ can be defined as the product of the Majorana mode
$\gamma_i$ and the fermion parity of the neighboring cloud of the conventional
electronic states $c_x$ which is unaffected by this process.  With these
macro-Majorana operators, the logical qubit states  can be defined by the
logical Pauli operators $\tilde{Z}=i\Gamma_{1}\Gamma_{2}=i\tilde P
\Gamma_{3}\Gamma_{4}$ and $\tilde X = i\Gamma_{2}\Gamma_{3}=i\tilde
P\Gamma_{1}\Gamma_{4}$ with $\tilde P = -\Gamma_1 \Gamma_2 \Gamma_3 \Gamma_4$
the total fermion parity of the system.  Then, the eigenstates associated with
these parity operators are robust quantum states as long as only the
interaction with the environment only generates localized quasi-particles.
Thus, the macro-Majorana proposal is particularly efficient to encode the
quantum memory into a topological vortex, as \emph{e.g.}, in
Ref.~\onlinecite{Sau}.  In this setup, one usually suffers from the presence
of an extremely small minigap, allowing for excitations at very low energies
thereby rendering the Majorana modes very fragile.  By introducing the
macro-Majorana operator $\Gamma_{i}$ encapsulating both the Majorana mode
$\gamma_{i}$ plus the surrounding cloud of excited states, the Majorana modes
$\Gamma_{i}$ become immune to localized excitations inside the vortex cores
thus solving the minigap problem.\cite{Akhmerov2010}

For the topological nanowire proposal we want to consider here, a similar
macro-Majorana encoding has not been analyzed so far.  The macro-Majorana
modes (take $\Gamma_{1}$ for example) are dephased only when the
quasi-particle after being excited close to $\gamma_1$ travels to the other
half of the nanowire, crossing from $S_1$ to $S_2$.  As long as the
quasi-particle remain localized, the quantum information encoded in the
macro-Majorana modes is intact.  If the quasi-particle on the other hand
crosses the virtual line, the logical $\tilde X$ flips resulting in a sign
flip error.
The naive guess for the coherence time $t_{\text{coh}}\approx
t_{\text{FGR}}+L/2v_{g}$ of the macro-Majorana encoding in a clean nanowire is
thus just the sum of the time $ t_{\text{FGR}}$ needed to excite the
quasi-particle above the superconducting gap (given by a Fermi golden rule)
plus the time $L/2v_{g}$ needed to travel the distance $L/2$ corresponding to
half the length of the wire; here, $v_g$ denotes the particles group velocity.
As $v_g \to 0$ when the energy of the quasi-particle approaches the
superconducting gap, one would expect $t_\text{coh} \approx L/2 v_g$,
\emph{i.e.}, proportional to the length of the wire $L$ with a possible large
prefactor due to the small $v_g$.  Moreover, including disorder in the wire
which renders the propagation of the quasi-particle diffusive or even localizes
the state and even longer coherence time $t_\text{coh}$ might be expected.

In this article, we address the question of the decoherence of a topological
superconductor wire hosting two Majorana modes at its boundaries.  We shall
show that the zero-energy modes are only protected by the presence of the gap.
In particular, the length of the wire does not help to obtain longer coherence
time of the quantum memory construction.  This is because---at least for a
clean system---the quasi-particles responsible for the decoherence of the
qubit propagate at a velocity of the order of the Fermi velocity $v_{F}$.  As
the Fermi velocity is usually rather large, the coherence time of a Majorana
wire memory is limited by the probability for an extra quasi-particle to be
excited above the gap.  We illustrate this idea in the case of the
macro-Majorana construction (see Fig.~\ref{fig:Q_memory_encoding}) for the
special case of thermal noise.

The specific setup for which we obtain our results is a system initially
prepared at zero temperature (no quasi-particles present).  We then study
excitations generated by coupling the system to an environment during the time
$t$.  The coherence time of the qubit $t_\text{coh}$ is at low temperatures
dominated by processes which involve the local excitation of a single
quasi-particle above the proximity induced gap in the nanowire.  We neglect
effects due to breaking up Cooper pairs as well as the generation of
quasi-particles in the bulk superconductor as this involves higher energy
excitations.  We study a toy model of a clean, single band, spinless, chiral
$p$-wave Bogoliubov-de Gennes Hamiltonian.  We calculate the Fermi golden rule
result for the creation of an extra excited quasi-particle above the
superconducting gap in section \ref{sec:Noisy-interaction-FGR}, for an
environment with a noise spectrum corresponding to thermal noise, Lorentzian
noise spectrum, or non-equilibrium noise due to the coupling to a nearby
quantum point contact.  We show that generically the excited quasi-particles
propagate at the Fermi velocity and that almost no effects of the group
velocity $v_{g}\ll v_{F}$ are visible, see section \ref{sec:Decay-clean}.  We
shortly discuss the effect of disorder in section \ref{sec:decay-dirty}.

\section{Model and hypothesis\label{sec:Model-and-hypothesis}}

Originally, Kitaev's model involve a $p$-wave superconductor.\cite{Kitaev2001}  This state is characterized by a spinless Cooper-pair
condensate, which satisfies Pauli exclusion principle thanks to the odd parity
symmetry of the gap.\cite{Volovik2003}  A chiral $p$-wave superconductor can
be emulated with a conventional ($s$-wave) superconductor with strong
spin-orbit effect and broken time-reversal symmetry.  Indeed, the spin-orbit
effect is known to lift the inversion symmetry constraint, allowing the
superconducting gap to possess both singlet and triplet components.\cite{gorkov_rashba.2001}  Additionally, breaking time-reversal symmetry will
destroy Kramers degeneracy and allows that the Majorana modes appear unpaired.\cite{sau:10,alicea:10}  Thus, the combination of strong spin-orbit plus
Zeeman effects in a conventional superconductor in the right parameter regime
implements an effective topological superconductor hosting Majorana modes at
its ends.\cite{lutchyn:10,Oreg2010,referee2:10}  In practice, the superconductivity is
induced by proximity effect to a strong spin-orbit semiconducting wire,
whereas the Zeeman effect is induced by applying a magnetic field along the
wire.\cite{Franz2013}

To simplify the calculations, we start with the simplest model exhibiting
Majorana modes: a spinless $p$-wave superconducting wire.  This model is
particularly useful in the clean case, when it is formally equivalent to the
experimental situation.\cite{alicea:12}  In this section, we
discuss the coupling between the zero-energy modes and the excited modes above
the gap due to the interaction with an environment.

A $p$-wave superconductor is described by the Bogoliubov-de Gennes (BdG)
Hamiltonian  in the so-called Andreev or quasi-classical
approximation,
\begin{equation}
H_{0}=v_{F}\hat{p}\sigma^{z}-v_{F}p_{F}\tau^{z}+\Delta_{x}\tau^{y}\sigma^{y}-\Delta_{y}\tau^{x}\sigma^{y}\label{eq:H0}
\end{equation}
where $\tfrac12v_{F}p_{F}=\mu_0$ denotes the chemical potential, the momentum
operator $\hat{p}=-i\hbar\partial_{x}$ in space representation, and the
complex superconducting gap $\Delta_{0}=\Delta_{x}+i\Delta_{y}$ ($\Delta_{x}$
and $\Delta_{y}$ are real) is supposed to be space-independent---hereafter we
denote $\Delta=\Delta_{0} e^{i\varphi}$, $\Delta_0>0$ and choose $\varphi=0$
because the phase of the superconducting order parameter is unimportant as we
have only a single superconductor in our setup and thus coherence effects are
absent.  The $\sigma^{i}$ and $\tau^{i}$ are Pauli matrices and act in the
propagating (right/left moving particles) and particle-hole spaces,
respectively. 

The BdG Hamiltonian Eq.~\eqref{eq:H0} exhibits a topologically phase with two
zero-energy modes located at the two ends of the
wire.\cite{Kitaev2001,Sengupta2001} In the situation when the wire is much
longer than the coherence length $L\gg\xi=\hbar v_{F}/\Delta$, the eigenstates
of the BdG Hamiltonian $H_{0}$ are approximately given by
\begin{equation}
\left\langle x|0\right\rangle =\sqrt{\dfrac{2}{\xi}}\left(\begin{array}{c}
e^{i\pi/4}\\
e^{-i\pi/4}
\end{array}\right)e^{-x/\xi}\sin (k_{F}x)\label{eq:zero_energy_state}
\end{equation}
for the zero-energy state located on the left of the wire, with $\hbar
k_{F}=p_{F}$ and $H_{0}\left|0\right\rangle =0$, and
\begin{equation}
\left\langle x|q\right\rangle =\sqrt{\dfrac{2}{L}}\left(\begin{array}{c}
-1\\
1
\end{array}\right)\sin(qx)\sin(k_{F}x)\label{eq:excited_states}
\end{equation}
for the quasi-particle at energies above the gap $\Delta$, satisfying the
relativistic dispersion relation $(\varepsilon/\Delta)^{2}-(\xi q)^{2}=1$.

Note that $\left|q\right\rangle $ is an \emph{approximate} eigenstate of
$H_{0}$ at energy $\varepsilon\approx\Delta$ with $H_{0}\left|q\right\rangle
=\Delta\left|q\right\rangle +\mathcal{O}(\ell^{-1})$, where we have used
$\ell=L/\xi \gg 1$ as a large parameter.\cite{klinovaja:12} The eigenstate
$\left|0\right\rangle $ is located at the left of the wire, whereas the
excited states $\left|q\right\rangle $ are fully delocalized along the wire.
The excited modes given above are written in a quasi-continuum fashion,
whereas the wire geometry would exhibit some discrete modes.  See
App.~\ref{sec:Appendix:-p-wave-wire} for more details, in particular for
the exact solutions of $H_{0}\left|q\right\rangle
=\varepsilon\left|q\right\rangle $ satisfying the boundary conditions
$\left\langle x=0|q\right\rangle =\left\langle x=L|q\right\rangle =0$ of a
finite-length wire.  In addition to the exact solution of the quasi-particle
state, we have also included the expression for the second unpaired Majorana
mode wave-function located at the right end of the wire with $x\approx L$ 
which we do not need for the following discussion.

Starting with the wire at zero-temperature, there are no quasi-particles
present and the system is characterized by the occupation of the Majorana zero
modes.  We prepare the system in a specific state of the two level system
spanned by the logical operators $\tilde Z $ and $\tilde X$.  Initializing the
system in a specific eigenstate of $\tilde X$ (\emph{e.g.}, the state to the
eigenvalue $+1$), and turning on the interaction with the environment it is
possible that a local interaction involving $\gamma_1$ generates a
quasi-particle located near $x\approx0$ at energy $\varepsilon \approx \Delta$ just above
the proximity-induced gap.  A qubit sign error happens as soon as this mobile
quasi-particle crosses from the region $S_1$ to $S_2$, see
Fig.~\ref{fig:Q_memory_encoding}. Alternative processes which dephase the
qubit are given by breaking up a Cooper-pair and one of the generated
particles crossing from $S_1$ to $S_2$ which involves at least an energy
$2\Delta$ and the generation of quasi-particles in the bulk superconductor
which are at even higher energies. Both of these processes are neglected in
the following as we want to concentrate on those processes which need the
least energy input from the environment and thus are dominant at very low
temperatures.

Let us discuss the possible interaction mechanisms of the environment with the
nanowire: in practice, the $p$-wave superconductivity is induced by proximity
of a strong spin-orbit semi-conductor with a conventional (non-topological)
superconductor.\cite{Oreg2010} The noise might originates from variations in
the applied magnetic field along the semiconductor wire generating
fluctuations in the induced Zeeman effect inside the wire, or even influencing
the proximity effect.  This latter effect may introduce fluctuations in the
induced gap parameter.  Possible other sources acting on the superconducting
gap are local magnetic impurities, or local Josephson vortices resulting from
imperfect deposition of the two materials during the sample preparation.  In
the following, we disregard these effects which lead to variations of the
superconducting order parameters as we believe that they are of minor
importance.  On the other hand, an imperfect contact between the
superconductor and the semiconductor and nearby fluctuating gates or mobile
charge impurities can lead to local fluctuations of the chemical potential.  A
time-dependent chemical potential $\mu(t) = \mu_0 + V(t) $ can be
incorporated in the model Hamiltonian \eqref{eq:H0} via
\begin{equation}
H=H_{0}+V(t)\tau^{z}\label{eq:H_V}
\end{equation}
with a generic time-dependent potential $V(t)$.

In the following, we need the interaction matrix element $M(q)=\left\langle
q\right|\tau^{z}\left|0\right\rangle $.  Evaluation in the limit of long wire
gives
\begin{equation}
M(q)=\sqrt{\dfrac{2}{\ell}}\dfrac{\xi q}{1+(\xi q)^{2}}\label{eq:interaction_element}
\end{equation}
as the probability amplitude for the zero-energy mode to scatter to
an excited state slightly above the gap. For convenience, we define
the wave-vector $\xi q=\sinh\eta$ and the energy $\varepsilon=\Delta\cosh\eta$
in term of the rapidity $\eta$, such that
\begin{equation}
M(q)=\sqrt{\dfrac{2}{\ell}}\dfrac{\sinh\eta}{\cosh^{2}\eta}
\end{equation}
in this parameterization.  The reparameterization has advantages when
manipulating the integrals of the following sections, since it makes the
relativistic dispersion relation of the quasi-particles explicit (see in
particular App.~\ref{sec:Integral-evaluation}).

It might be unclear whether Eq.~(\ref{eq:interaction_element}) represents
or not the genuine matrix element coupling the states $\left|0\right\rangle $
and $\left|q\right\rangle $. This is because the excited states $\left|q\right\rangle $
are \emph{not exact} eigenstates of $H_{0}$. In particular, using
the notations of Eqs.~(\ref{eq:zero_energy_state},\ref{eq:excited_states}),
we easily find that $\left\langle q|0\right\rangle \propto\left\langle
q\right|\tau^{z}\left|0\right\rangle \neq0$.
The \emph{exact} excited states found in the App.~\ref{sub:Finite-wire}
are nevertheless orthogonal to the zero-energy mode $\left|0\right\rangle $,
and the interaction element can be shown to be exactly the one above
in the long wire limit $\ell\rightarrow\infty$. More explicitly,
one can show that $\left\langle 0|q\right\rangle \propto e^{-\ell}$
whereas $\left\langle 0\right|\tau^{z}\left|q\right\rangle \propto\ell^{-1/2}$
as in Eq.~(\ref{eq:interaction_element}), using the exact excited
states $\left|q\right\rangle $ found in the App.~\ref{sub:Finite-wire}.
To remedy the use of the approximate excited states (\ref{eq:excited_states})
in the following calculations, we will keep the $\tau^{z}$ matrix,
and use the \emph{exact} algebra $\left\langle 0\right|\tau^{z}\left|0\right\rangle =0$,
$\left\langle q\right|\tau^{z}\left|0\right\rangle =M(q)$
and $\left\langle q|0\right\rangle =0$.

We note that the interaction element $M(q)$ does not couple the zero-energy
mode to the mode exactly at the energy gap  (corresponding to $\eta=0$ in our
parameterization).  This helps for the stability of the quantum memory since
the density of state $\rho=\partial q/\partial\varepsilon=\left(\hbar
v_{F}\tanh\eta\right)^{-1}$ diverges at the gap.\cite{b.tinkham}

\section{Interaction with the environment: a Fermi golden rule
approach for the local Majorana encoding\label{sec:Noisy-interaction-FGR}}

In this section, we study the evolution operator associated to our model
Hamiltonian $(\ref{eq:H_V})$ in order to obtain the probability transition of
the zero-energy mode to the quasi-continuum, according to the Fermi golden
rule.\cite{Schoelkopf2002} Note that the Fermi golden rule gives the coherence
time $t_\text{coh}$ of the local qubit encoding with $\gamma_i$ but not of the
macro-Majorana encoding with $\Gamma_i$ as it does not take into account the
time it takes for the excited quasi-particle to travel the distance $L/2$.  The
Fermi golden rule is the relevant result if one can suppose instantaneous
propagation along the wire or when the quantum memory is encoded in
terms of the local Majorana modes $\gamma_{i}$ instead of the macro-Majorana
$\Gamma_{i}$.\cite{Schmidt2012} We will first start with the results using the
Fermi golden rule approach before we will introduce the effects of the
propagation in the following section.

First, we suppose that the interaction potential is so weak that the
truncation at first order of the evolution operator
\begin{equation}
U(t)\approx
U_{0}(t)+\dfrac{1}{i\hbar}\int_{0}^{t}U{}_{0}^{\dagger}(\tau)V(\tau)\tau^{z}U_{0}(\tau)d\tau\label{eq:Unitary_evolution_app}
\end{equation}
is valid, with $U_{0}(t)=e^{-itH_{0}/\hbar}$. 

Then, we define the noise spectrum $S(\omega)$ in term
of the interaction potential as
\begin{equation}
\left\langle V(t_{1})V(t_{2})\right\rangle_{\text{noise}}=\int\dfrac{d\omega}{2\pi}e^{i\omega(t_{2}-t_{1})}S(\omega)\label{eq:noise_spectrum}
\end{equation}
where the average is over all configurations of the noise.\cite{Schoelkopf2002}
We also assume that $\left\langle V(t)\right\rangle_{\text{noise}}=0$ as a
nonzero average simply leads to a redefinition of the chemical potential
$\mu_0$.

The probability $P_\gamma(t)$ to excite a zero-energy mode
$\left|0\right\rangle $ to an arbitrary state in the quasi-continuum states
$\left|q\right\rangle $ is defined as
\begin{align}
P_\gamma(t) & =\int\dfrac{Ldq}{\pi}\left\langle \left|\left\langle q\right|U(t)\left|0\right\rangle \right|^{2}\right\rangle _{\text{noise}}\nonumber \\
 & =\dfrac{1}{\hbar^{2}}\int\dfrac{Ldq}{\pi}\int\dfrac{d\omega}{2\pi}S(\omega)\left|g_{\text{FGR}}(\omega,t)\right|^{2}
\end{align}
with 
\begin{align}
g_{\text{FGR}}(\omega,t) & =\int_{-t/2}^{t/2}d\tau
e^{-i\omega\tau}\left\langle
q\right|e^{\frac{i\tau}{\hbar}H_{0}}\tau^{z}\left|0\right\rangle .
\end{align}
For large time we can replace $\left|g_{\text{FGR}}(\omega,t)\right|^{2}$ by
$2\pi t\delta\left(\omega-\omega_{\Delta}\cosh\eta\right) \left|M(q)
\right|^{2}$,\cite{note1} where we have introduced $\hbar\omega_{\Delta} =
\Delta$ and neglected the contribution $\left\langle q|0\right\rangle \propto
e^{-\ell}$ valid in the limit of large $\ell$.

The probability per unit time for a zero-energy mode to get excited in any
state of energy above the energy gap is given by
$\Gamma_{\text{FGR}}=dP_\gamma/dt$ where
\begin{align}
\Gamma_{\text{FGR}} &
=\dfrac{2}{\pi\hbar^{2}}\int_{0}^{\infty}\dfrac{\sinh^{2}\eta}{\cosh^{3}\eta}S\left(\omega_{\Delta}\cosh\eta\right)d\eta\label{eq:Gamma_FGR};
\end{align}
a result known as the Fermi golden rule.\cite{Schoelkopf2002} 

We are interested in the three particular forms of noise spectrum
\begin{equation}
S(\omega)=\begin{cases}
S_{0}\exp\left[-\hbar\omega/k_{B}T\right], & \text{thermal,}\\
S_{0}\left[1+(\omega-\omega_{0})^{2}/\alpha^{2}\right]^{-1}, &
\text{Lorentzian,}\\
S_{0}(1-\hbar\beta\omega)\;;\;\hbar\omega\beta\leq1, & \text{QPC,}
\end{cases}\label{eq:noise_spectrum_definitions}
\end{equation}
with $S_{0}$ a characteristic amplitude for the noise spectrum.  The first
line corresponds to the equilibrium noise spectrum for a contact with a bath
at temperature $T$.  The second line of
$(\ref{eq:noise_spectrum_definitions})$ corresponds to the case of a
Lorentzian shape noise power with a center frequency $\omega_0$ and a
bandwidth $\alpha$.  In the Lorentzian model, the transition between a
quasi-monochromatic noise spectrum when $\alpha\rightarrow0$ and a
quasi-white-noise with all frequencies equally excited when
$\alpha\rightarrow\infty$ can be described.  The last model we discuss is the
case of the excess noise of a quantum point contact (QPC).  In that case,
$S_{0}=\sum_{n}T_{n}(1-T_{n})e^{3}V/\pi\hbar$ with $T_{n}$ the $n$-th
transmission eigenvalue of the barrier between the wire and an electronic
reservoir at zero-temperature, $V$ the voltage drop of the barrier, and
$\beta=1/eV$ (see \emph{e.g.}, Refs.~\onlinecite{Blanter2000}).

We start with thermal noise.  Since we are interested in the regime when the
Majorana modes are well defined, we focus on the low temperatures regime
$T\ll\Delta$ as otherwise quasi-particle destroying the quantum memory are
ubiquitous; see Ref.~\onlinecite{Catelani2012} for a more general discussion
of the superconducting qubit systems.  In the low temperatures limit, the
integral in \eqref{eq:Gamma_FGR} is dominated at small wave-vectors and we
obtain
\begin{align}
\Gamma_{\text{FGR}} & \approx\dfrac{2S_{0}}{\pi\hbar^{2}}e^{-\frac{\Delta}{k_{B}T}}\int_{0}^{\infty}z^{2}e^{-\frac{z^{2}}{2}\frac{\Delta}{k_{B}T}}dz\nonumber \\
 & =\sqrt{\dfrac{2}{\pi}}\dfrac{S_{0}}{\hbar^{2}}\left(\dfrac{k_{B}T}{\Delta}\right)^{3/2}e^{-\frac{\Delta}{k_{B}T}}\label{eq:FermiGR_thermal_noise}
\end{align}
as found in the appendix of Ref.~\onlinecite{Schmidt2012}. The opposite
(experimentally not relevant) limit
$\Delta/k_{B}T\ll1$ gives a logarithmic correction
\begin{equation}
\Gamma_{\text{FGR}}\approx\dfrac{2S_{0}}{\pi\hbar^{2}}\left[\dfrac{\pi}{4}-\dfrac{\Delta}{k_{B}T}\ln\dfrac{\Delta}{k_{B}T}\right]
\end{equation}
of the decay rate. 

Next, we discuss the Lorentzian noise model.  The Fermi golden rule associated
to the Lorentzian spectral density can be calculated exactly, and gives
\begin{multline}
\dfrac{\pi\hbar^{2}\Gamma_{\text{FGR}}}{2S_{0}}=\dfrac{\alpha}{\omega_{\Delta}}\Im\left\{ \dfrac{1}{z_{0}^{2}}+\dfrac{\pi}{4z_{0}^{2}}\left(2-z_{0}^{2}\right)\right.\\\left.-\dfrac{2}{z_{0}^{2}}\sqrt{z_{0}^{2}-1}\arctanh\sqrt{\dfrac{z_{0}+1}{z_{0}-1}}\right\}     \label{eq:FermiGR_Lorentz}
\end{multline}
with $z_{0}=(\omega_{0}+i\alpha)/\omega_{\Delta}$.
\begin{figure}
\centering
\includegraphics[width=0.8\linewidth]{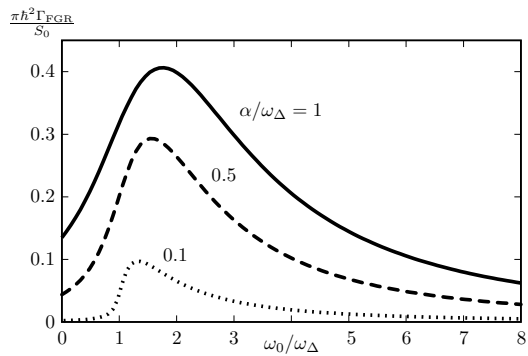}
\caption{Decay rate $\Gamma_{\text{FGR}}$ of the qubit encoded in the zero
modes of a topological superconducting wire in an environment having a
Lorentzian noise spectrum, Eq.~(\ref{eq:FermiGR_Lorentz}), as a function of
the center frequency $\omega_{0}$.  The different curves correspond to
different values of the broadening $\alpha/\omega_\Delta = 0.1,0.5,1$ as
indicated in the plot.  The condition $\omega_{0}=\omega_{\Delta}$ corresponds
to a Lorentzian noise spectrum with its maximum amplitude at the energy gap.
\label{fig:Fermi_golden_rule_Lorentz}}
\end{figure}

The decay time $\Gamma_{\text{FGR}}$ is plotted on
Fig.~\ref{fig:Fermi_golden_rule_Lorentz} for different values of $\alpha$ with
respect to the resonance frequency $\omega_{0}$ of the noise spectrum.  The
superconducting gap is well-visible in this plot. For small enough $\alpha$,
\emph{i.e.}, for quasi-monochromatic noise, the decay time
$\Gamma_{\text{FGR}}$ is negligible as long as the noise resonance frequency
$\omega_{0}$ is smaller than the frequency associated with the superconducting
gap $\omega_{\Delta}$, and then it has a peak a little bit above the
$\omega_{0}/\omega_{\Delta}=1$ angular frequency.  It then decays first
exponentially when $\omega_{0}>\omega_{\Delta}$, then as a power law for
$\omega_{0}/\omega_{\Delta}\gg1$.  For broader spectrum, the decay rate no
longer vanishes for frequencies below the gap, but rather it becomes more flat
over larger frequencies: the gap frequency is no more a characteristic
frequency since one can pump a lot of frequencies with approximately the same
amplitude.  For even larger bandwidths, \emph{i.e.}, in the white noise limit,
one pumps all the frequencies at an approximately equal amplitude, so the amplitude
to switch to any high-energy level is almost flat.  It is noteworthy that a broad enough noise spectrum can by
itself poisons the system with quasi-particles.  We believe this poisoning is
not intimately related to our topological model for the superconducting wire,
and may be a more general issue valid for any kind of superconducting system.
Of course, our model predicts the first excited states to be at energy
$\Delta$ since the zero-energy mode is populated in our system, whereas the
conventional superconductivity would have an excitation energy above
$2\Delta$.

\begin{figure}
\centering
\includegraphics[width=0.8\linewidth]{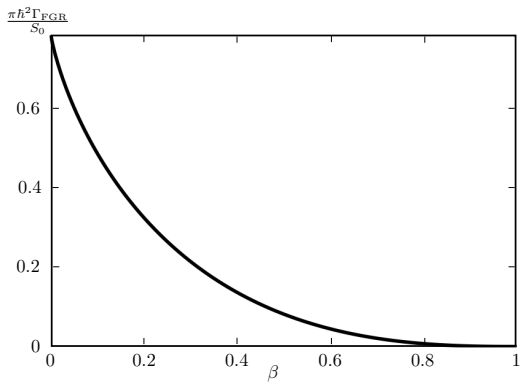}
\caption{Fermi golden rule decay rate $\Gamma_{\text{FGR}}$ of the quantum
memory in proximity to a quantum point contact, Eq.~(\ref{eq:FermiGR_QPC}), as
a function of $\beta=(eV)^{-1}$ representing the inverse voltage drop across
the contact.
\label{fig:Fermi_golden_rule_QPC}}
\end{figure}

Finally, the Fermi golden rule for a  QPC
leads to the result
\begin{multline}
\dfrac{\pi\hbar^{2}\Gamma_{\text{FGR}}}{4S_{0}}=\arctan\dfrac{1+\sqrt{1-\beta^{2}}}{\beta}-\dfrac{\beta}{2}\ln\dfrac{\sqrt{1-\beta^{2}}+1}{\sqrt{1-\beta^{2}}-1}\\
+\dfrac{\beta}{2}\sqrt{1-\beta^{2}}-\dfrac{\pi}{4}\label{eq:FermiGR_QPC}
\end{multline}
which is plotted in Fig.~\ref{fig:Fermi_golden_rule_QPC}
as a function of $\beta$. For large voltage difference between the wire
and the environment (small $\beta$), the transition amplitude is
high, then decay and goes algebraically to zero for smaller voltages.
When $eV\geq1$, the barrier is no more transmitting, and the excitation
probability $\Gamma$ goes to zero.

\section{Propagation along the wire and decoherence in the macro-Majorana
encoding\label{sec:Decay-clean}}

\begin{figure}
\centering
\includegraphics[width=0.9\linewidth]{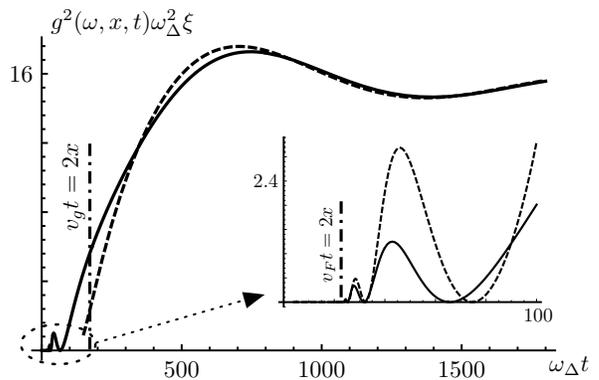}
\caption{Comparison between the asymptotic evaluation of $g^{2}(\omega,x,t)$
and the exact numerical results for a given $\omega$ and $x$, as a function of
time $t$ for $x/\xi=12$ and $\omega/\omega_{\Delta}=1.01$.  The plots
represent the probability distribution for a quasi-particle excited at
an energy $\hbar \omega$ to reach the point $x$ in the time $t$.  We compare
the exact result (solid line) with the asymptotic expansions
Eq.~(\ref{eq:long-time}) (dashed curve in the main panel) and
Eq.~\eqref{eq:intermediary_times} (dashed curve in the inset).  We have
indicated the two relevant time-scale $2x/v_g$ corresponding to the group
velocity and $2x/v_F$ corresponding to the Fermi velocity.}\label{fig:g_omega}
\end{figure}

In this section, we evaluate the probability for a quasi-particle
to be excited by an environment \emph{and} to propagate to the
second half of a clean wire. This mechanism is responsible for a qubit-flip,
then destroying the quantum memory in the macro-Majorana encoding
of Fig.~\ref{fig:Q_memory_encoding}. Our goal is to calculate
the expression
\begin{equation}
P_{\Gamma}(t)=\int_{L/2}^{L}\left\langle \left|\left\langle x\right|U(t)\left|0\right\rangle \right|^{2}\right\rangle _{\text{noise}}dx\label{eq:P}
\end{equation}
which is the macro-Majorana equivalent of the corresponding expression
$P_\gamma(t)$ for the local encoding.  We will show that the excited
wave-packet propagates at an effective velocity close to the Fermi velocity.
This section also shows how the Fermi golden rule is recovered when more
microscopic details are taken into account.  Indeed, we will explain that the
Fermi golden rule is a valid result at intermediate times (at infinite times,
the probability saturates, at small times it goes like
$t^2$).\cite{tannoudji,hassler:10} Although the excited quasi-particle should
propagate at a group velocity corresponding to the energy $\hbar\omega$
(Fig.~\ref{fig:g_omega}), we will find that the vanishing of the matrix
element $M(q)$ close to the gap only allows excitation of quasi-particles
whose group velocity essentially is given by the Fermi velocity.

Starting from Eq.~\eqref{eq:P}, we arrive after some algebra at 
\begin{equation}
P_{\Gamma}(t)=\dfrac{1}{\hbar^{2}}\int_{L/2}^{L}dx\int\dfrac{d\omega}{2\pi}S(\omega)g^{2}(\omega,x,t)\label{eq:P_S}
\end{equation}
with\cite{note2}
\begin{equation}
g(\omega,x,t)=\int_{-t/2}^{t/2}d\tau\left[e^{-i\omega\tau}\left\langle
x\right|e^{iH_{0}\tau/\hbar}\tau^{z}\left|0\right\rangle \right]\label{eq:g_clean}
\end{equation}
a generalization of $g_\text{FGR}$ of the last section.  The evaluation of
\eqref{eq:g_clean} is rather involved and we have moved the details to
App.~\ref{sec:Integral-evaluation}.
As a result, the two following asymptotic regimes are found written with the
dimensionless variables $\tilde{x}=x/\xi$, $\tilde{t}=\omega_{\Delta}t$ and
$\tilde{\omega}=\omega/\omega_{\Delta}$, representing position and time, and
the noise spectrum frequency rescaled by the superconducting characteristic
length and frequency, respectively:

\begin{multline}
g(\left(\omega-\omega_{\Delta})t\gg1\right)\omega_{\Delta}\approx\dfrac{4}{\sqrt{\xi}}\left[\dfrac{\sin\left(\tilde{x}\sqrt{\tilde{\omega}^{2}-1}\right)}{\tilde{\omega}}\right.\\
\left.-\dfrac{4}{\sqrt{\pi}}\dfrac{\tilde{x}}{\tilde{t}^{3/2}}\dfrac{\sin\left(\left(\tilde{\omega}-1\right)\dfrac{\tilde{t}}{2}+\dfrac{\pi}{4}\right)}{\tilde{\omega}-1}\right]\label{eq:long-time}
\end{multline}
for large time $(\omega-\omega_{\Delta})t\gg1$ and
\begin{multline}
g(v_{F}t\gg x  \gg v_g t)\omega_{\Delta}
\!\approx\!
\dfrac{8}{\sqrt{\pi\xi}}\dfrac{\tilde{x}}{\tilde{t}^{3/2}}\dfrac{\sqrt{(\tilde{t}/2)^{2}-\tilde{x}^{2}}}{\tilde{\omega}\sqrt{(\tilde{t}/2)^{2}-\tilde{x}^{2}}-\tilde{t}/2}\\
\times\cos\left(\sqrt{\left(\tilde{t}/2\right)^{2}-\tilde{x}^{2}}-\tilde{\omega}\dfrac{\tilde{t}}{2}+\dfrac{\pi}{4}\right)\label{eq:intermediary_times}
\end{multline}
when $v_{F}t\gg x$ and $\omega x\gg
v_{F}t\sqrt{\omega^{2}-\omega_{\Delta}^{2}}$.  The velocity
$v_{g}=v_{F}\sqrt{1-\omega_{\Delta}^{2}/\omega^2}$ represents a group
velocity corresponding to the excitation at the frequency $\omega$.  Note that
for $\omega \gtrsim 1$, we have $v_g \ll v_F$ whereas for $\omega \gg 1$ $v_g
\approx v_F$.
\begin{figure} 
  \centering 
  \includegraphics[width=0.7\linewidth]{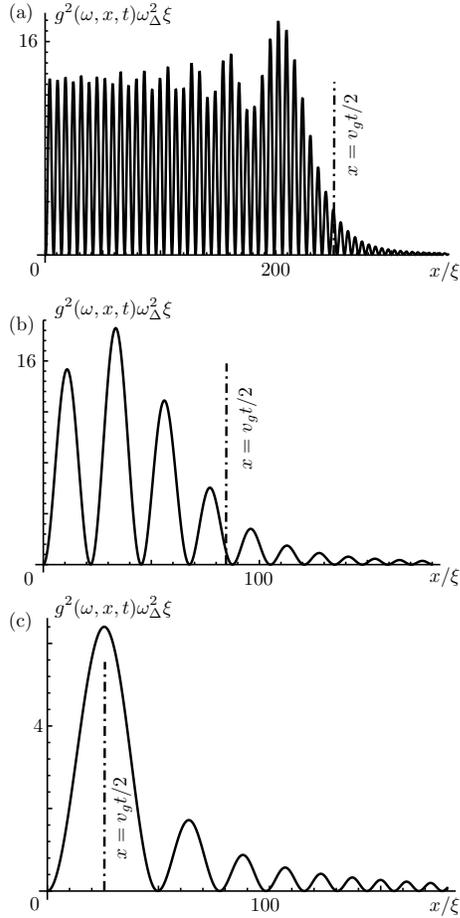}
\caption{Plot of $g^{2}(\omega,x,t)$ as a function of $x$ with
$v_{F}t=1200\xi$ for different frequencies
$\tilde{\omega}=\omega/\omega_{\Delta}$.  (a): $\tilde{\omega}=1.1$,
$v_{g}t/2\approx500\xi$ ; (b): $\tilde{\omega}=1.01$, $v_{g}t/2\approx85\xi$;
(c): $\tilde{\omega}=1.001$, $v_{g}t/2\approx25\xi$.  The two first plots are
well approximated by a sine function, according to Eq.~(\ref{eq:long-time})
up to the group velocity $v_{g}t/2$; see the discussion below
Eq.~(\ref{eq:intermediary_times}).  The position $v_gt/2$ is represented by
the dot-dashed vertical line in each plot.  The bottom plot (c) corresponds to
a frequency very close to the superconducting gap, when
Eq.~(\ref{eq:long-time}) is not  valid.\label{fig:g_omega-X}}
\end{figure}

We discuss the results of $g(\omega, x,t)$ for a specific position $x$ and
frequency $\omega$ as a function of $t$, see Fig.~\ref{fig:g_omega}:
initially, \emph{i.e.}, for $t \leq x/v_F$, no quasi-particle propagation has taken
place and $g(\omega,x,t) \approx 0$.  At intermediary times $x/v_F \leq t\leq
x/v_g$ equation~(\ref{eq:intermediary_times}) is valid and the probability
density to have an extra quasi-particle at position $x$ oscillates and grows
as time is passing up to an (apparent) divergence at the group velocity
$v_{g}$.  For large times $t \geq x/v_g$ equation~(\ref{eq:long-time}) is
valid and the probability density saturates to a finite value, establishing a
totally delocalized quasi-particle probability distribution (first term of
Eq.~(\ref{eq:long-time})) when $t\rightarrow\infty$.  In between these two
regimes, there is a monotonous increase of the probability amplitude which we
will discuss in more details below.

Alternatively, we can understand the function $g^2(\omega,x,t)$ at a fixed
time $t$ as a function of position $x$, see Fig.~\ref{fig:g_omega-X}.  For not
too small energies $\hbar \omega$, Fig.~\ref{fig:g_omega-X} (a) and (b), the
main part of quasi-particle probability distribution is situated at $x \leq
v_{g}t/2$ where the result \eqref{eq:long-time} is valid.  This suggests to
approximate $g^2(\omega, x ,t )$ as
\begin{equation}\label{eq:approx}
  g^2(\omega,x ,t) \approx  \frac{16 \sin^2(\tilde x \sqrt{\tilde \omega^2
  -1})}{\omega^2_\Delta \xi \tilde\omega^2}  \Theta(t v_g/2 -x),
\end{equation}
\emph{i.e.}, using the first contribution of Eq.~(\ref{eq:long-time}) which
encapsulates the position and frequency dependency of the quasi-particles
delocalization, and to neglect space contributions above distance $v_{g}t/2$;
here, $\Theta(x)$ denotes the unit-step function.

As a consistency check of this approximation, let us now discuss how to
recover the Fermi golden rule (\ref{eq:Gamma_FGR}) in the long time limit.
Since the Fermi golden rule does not take into account the space dependency of
the probability, we have to define $P_{\gamma}(t)=\int_{0}^{L}\left\langle
\left|\left\langle x\right|U(t)\left|0\right\rangle \right|^{2}\right\rangle
_{\text{noise}}dx$ as the probability for an extra quasi-particle to be found
anywhere in the wire.  In comparison with the definition (\ref{eq:P})
corresponding to the macro-Majorana encoding, $P_{\gamma}$ corresponds to the
local Majorana encoding.  Using the result
Eq.~(\ref{eq:approx}), we obtain
\begin{equation}
P_{\gamma}(t)\approx\dfrac{8}{\pi\hbar^{2}}\int
d\omega\dfrac{S(\omega)}{\omega^{2}}\int_{0}^{\frac{v_{g}\tilde{t}}{2v_{F}}}\sin^{2}\left(\tilde{x}\sqrt{\tilde{\omega}^{2}-1}\right)d\tilde{x}.\label{eq:P_gamma}
\end{equation}
For $v_{g}\tilde{t}/v_{F}\gg\left(\tilde{\omega}^{2}-1\right)^{-1/2}$, the
sine function in the integral can be approximated by its mean value
$\sin^{2}\left(\tilde{x}\sqrt{\tilde{\omega}^{2}-1}\right)\approx1/2$, and we
end up with exactly the Fermi golden rule (\ref{eq:Gamma_FGR}), provided we
use the definitions $\omega=\omega_{\Delta}\cosh\eta$ and
$v_{g}=v_{F}\tanh\eta$.
\begin{figure}
\centering
\includegraphics[width=0.95\linewidth]{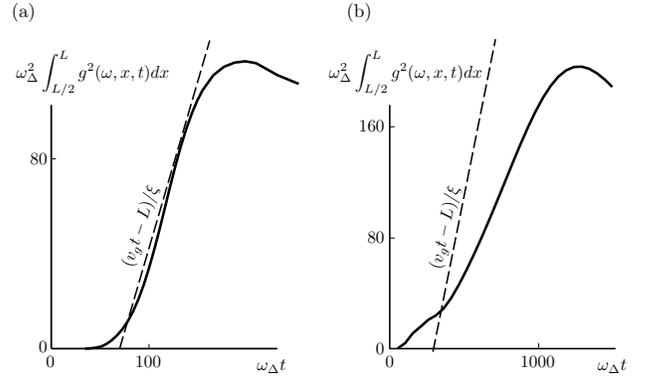}
\caption{Integrated probability amplitude $\int_{L/2}^{L}g^{2}(\omega,x,t)dx$
as a function of time $t$ for a fixed length of the wire $L=40\xi$ and for an
environment exciting at frequencies $\omega=1.2\omega_{\Delta}$ (left panel),
and $\omega=1.01\omega_\Delta$ (right panel). A comparison with the
linear slope of $(v_{g}t-L)/\xi$ of Eq.~(\ref{eq:P-Gamma-L}) is provided.  The
approximation is rather good for large frequencies (left panel) whereas it
fails for small frequencies (right panel).\label{fig:Xint}}
\end{figure}

\begin{figure}
\centering
\includegraphics[width=0.8\linewidth]{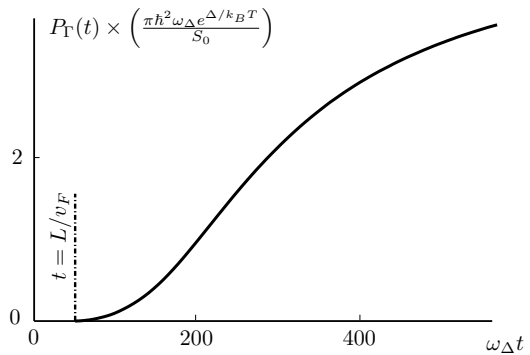}
\caption{Plot of $P_\Gamma(t)$ in units of $S_0\exp(-\Delta/k_{B}T) /
\pi\hbar^{2}\omega_{\Delta}$ from Eq.~(\ref{eq:P}), with respect to time
$\tilde{t}=\omega_{\Delta}t$, for a thermal noise, when $\Delta/k_{B}T=20$ and
$\ell=50$.  We also represented the characteristic time $L/v_{F}$
corresponding to the propagation with the Fermi velocity $v_F$ see the
discussion in the main text.\label{fig:P-Gamma}}
\end{figure}

Returning to the task to evaluate $P_{\Gamma}$ using the approximation
\eqref{eq:approx} for $g^2(\omega,x,t)$.  The result is (neglecting fast
oscillatory terms)
\begin{multline}
P_{\Gamma}(t)\approx\dfrac{4}{\pi\hbar^{2} \xi}\int
d\omega\dfrac{S(\omega)}{\omega^{2}}\left(
v_g t-L\right) \Theta\left(v_g t-L\right)\label{eq:P-Gamma-L}
\end{multline}
for a comparison of the exact result with the approximation given above see
Fig.~\ref{fig:Xint}.  This result seems to indicates that the qubit start to
dephase at a characteristic time $L/v_g$.  Note however that last equation is
only correct away from the regime with $\omega\approx\omega_{\Delta}$ where
$v_g \approx v_F$ since the approximation (\ref{eq:approx}) is not valid in
this limit.  In fact for energies close to the gap with $\omega \approx
\omega_\Delta$ there is no sharp feature visible in $P_\Gamma$ associated with
the group velocity instead $P_\Gamma$ monotonously grows starting at a time
$L/v_F$.  In App.~\ref{sec:Integral-evaluation} this result is associated to
the fact that the saddle point giving the contribution at $L/v_g$ becomes
broad right in the regime where $v_g \ll v_F$.  In conclusion, we find that
there is only a sharp feature at the group velocity visible in the case where
$v_F \approx v_g$ and for the case $v_g \ll v_F$ where we would expect an
increase of the coherence of the quantum memory due to the slow motion of the
quasi-particle the corresponding feature is washed out.  As an example, we
have numerically calculate $P_\Gamma$ for a thermal environment in
Fig.~\ref{fig:P-Gamma} at temperature $\Delta = 20 k_B T$.  From the plot, it
is clear that the characteristic time for the decoherence of the quantum
memory is given $L/v_F$ which corresponds to a characteristic speed $v_F$ of
the involved quasi-particles even though in a naive picture only particles
close to the gap with $v_g \ll v_F$ are excited.  The example of the thermal
noise shows that we can approximate coherence time of the macro-Majorana
encoding as $t_{\text{coh}}\approx t_{\text{FGR}}+L/v_{F}$ even for low
temperatures.  We estimate a propagation time $L/v_{F}\simeq
1\,\text{fs}$ for a wire with a length of a few micrometers lengthand
$v_{F}\simeq 10^{8}\,\text{cm}/\text{s}$.\cite{mourik:12} For experiments at
sufficiently low temperatures, we expect $ t_{\text{FGR}} \gg L/v_{F}$ and
thus the macro-Majorana encoding will generically not provide a better
stability than the local encoding via $\gamma_i$. Concluding, the quantum
memory encoded in the Majorana modes is only protected due to the gap.  In
particular, any kind of \emph{local} interaction at frequencies $\omega \geq
\omega_\Delta$ in the proximity of the location of the Majorana mode is
sufficient to immediately (up to a small correction of magnitude $L/v_F$)
destroy the quantum memory.

\section{Discussion: does disorder help to localize the quasi-particles?\label{sec:decay-dirty}}

Since we have found that the length of the wire does not increase the
coherence time of the quantum memory in the clean limit studied so far, one
might wonder if disorder which decreases the speed of propagation of the
quasi-particles might help to increase the coherence time.  For the toric code
in 2D, it has been shown that disorder helps to localize the quasiparticles
and thus increases the storage time of the quantum
memory\cite{Stark2011,Wootton2011} and similar results have been obtained for
a 1D setting very similar to the one studied here \cite{Bravyi2011}.

It is quite clear that it is not possible to enter the regime where the motion
of the quasiparticles is diffusive or where they are even localized as
$p$-wave superconductivity is known to be fragile to
impurities.\cite{b.benneman_ketterson,brouwer:11} Indeed, $p$-wave
superconductivity has only a particle-hole symmetry, in contrast to the
conventional $s$-wave superconductivity, which also is time-reversal symmetric
and is therefore immune to non-magnetic impurities.  Thus, having a Majorana
wire which is strongly disordered does not help increasing the robustness of
the Majorana mode wire encoding as the $p$-wave proximity effect is suppressed
when increasing the disorder strength.  The case of a moderate disorder
requires more careful attention.  As the physics is not universal in this
case, it is necessary to study a more realistic model of the nanowire
including multiple modes, $s$-wave pairing, spin-orbit, and a Zeeman field in
this case.\cite{Oreg2010,brouwer:11,referee-2-2} For the
moderately dirty system, a quasi-classical approach superconducting transport
in the form of the Eilenberger equations can be employed which can be
perturbatively expanded for a small amount of
impurities.\cite{kogan.1985,Neven2013} A study using the quasi-two-dimensional
version of the $s$-wave Eilenberger equation in the presence of strong
spin-orbit effect, moderate Zeeman interaction and few amount of disorder is
outside the scope of the present manuscript and thus postponed for further
studies.

\section{Conclusion}

We have discussed in details the interaction of a clean topological
superconductor wire with an environment, with the particular emphasis on the
propagation of the excited quasi-particles above the energy gap.  The
propagation of the quasi-particles becomes important when one considers the
macro-Majorana encoding of the quantum memory.  In particular, one would
expect that in this encoding a longer wire would increase the coherence time
of the memory.  Calculating the coherence time using the system-environment
coupling as a perturbation, we found that the quasi-particle excitations
generically propagate at the Fermi velocity (section \ref{sec:Decay-clean})
and that no sharp feature associated with a possible slower group velocity is
present.  As the Fermi velocity is typically rather large, this result implies
that the macro-Majorana encoding is not more robust than the local encoding
for the case of 1D nanowires.  In particular, we have shown that the
probability to excite the zero-energy mode into excited quasi-particle above
the superconducting gap is the principal mechanism of decay of the quantum
memory encoded in a Majorana clean wire.  This work puts strong constraints on
the usefulness of Majorana fermions as a quantum memory as the coherence time
is only dictated by the size of the gap without an additional benefit due to
the length of the wire.

\begin{acknowledgments} We thank Gianluigi Catelani for some insightful
remarks during the work, and Manuel Schmidt for discussing the result of
Ref.~\onlinecite{Schmidt2012} with us.  We also thank Barbara Terhal,
Christoph Ohm, Jascha Ulrich, and Giovanni Viola for stimulating discussions.
We are grateful for support from the Alexander von Humboldt foundation.
\end{acknowledgments}

\appendix

\section{$p$-wave superconducting wire of length $L$\label{sec:Appendix:-p-wave-wire}}

In this appendix, we discuss the general solutions of the Bogoliubov-de
Gennes Hamiltonian associated with a superconducting system in the $p$-wave
state.  Then we calculate the midgap states associated with the boundary of a
semi-infinite $p$-wave wire in contact with a topologically trivial vacuum of
particle.  Finally, we discuss the finite length wire system embedded in
vacuum.  We calculate both the zero-energy Majorana modes located at the two
ends of the wire, in addition to the full spectrum of excited quasi-particles
states.  We also discuss shortly the limiting case of a wire longer than the
superconducting coherence length, the regime studied in the main paper.

The model at hand is a spinless $p$-wave Bogoliubov-de Gennes (BdG)
Hamiltonian 
\begin{equation}
H_{\text{BdG}}=\left(\dfrac{p^{2}}{2m}-\mu_0\right)\tau^{z}+\Delta_{x}\dfrac{p}{p_{F}}\tau^{x}-\Delta_{y}\dfrac{p}{p_{F}}\tau^{y}
\end{equation}
with $p$ the momentum, $m$ the effective mass of the electrons,
$\mu_0=p_{F}^{2}/2m$ the chemical potential, $p_{F}$ the Fermi momentum, and
$\Delta_{0}=\Delta_{x}+i\Delta_{y}$ the superconducting order parameter.  The
Pauli matrices $\tau^{i}$ act on the particle-hole space, and are useful to
describe the symmetries of the system.  The $p$-wave model is used for exotic
phase of superfluid and superconductors (see
Refs.~\onlinecite{Volovik2003,b.mineev_samokhin} for instance), and exhibits
two topological sectors in one dimension, being in the class D, with only a
particle-hole ($\mathcal{P}$-type) symmetry $\mathcal{P}=\mathcal{K}\tau^{x}$
such that $\left\{ H,\mathcal{P}\right\} =0$ with $\mathcal{K}$ the complex
conjugation operator, see \emph{e.g.}, Refs.~\onlinecite{Ryu2010,Fulga2011a}.

When the gap is small with respect to the Fermi level $\Delta\ll \mu_0$,
one can describe the physics of superconductivity in the linear approximation
of the spectrum close to the Fermi level, the so-called Andreev or
quasi-classic approximation. It supposes the chemical potential $\mu_0=v_{F}p_{F}/2$
to be the Fermi level, and then to expand the band structure $(p^{2}-p_{F}^{2})$
close to the Fermi level, where we must specify the direction of propagation.
We can in practice do that in a $4\times4$ Hamiltonian constructed
from $H_{\text{BdG}}$. Then, the linearization approximation also
requires to specify the direction of propagation $p/p_{F}\approx\text{sgn}(p)$
for the gap. We then obtain $H_{0}$ of the full text, see Eq.~(\ref{eq:H0}).
Note that due to the projection in the propagation basis, both $H_{\text{BdG}}$
and $H_{0}$ have the same $\mathcal{P}$-type symmetry with
$\mathcal{P}=\mathcal{K}\tau^{x}$.
In short, the Andreev approximation must preserve the topological
classification, at the expense of doubling the number of degrees
of freedom. This doubling in the degrees of freedom is nevertheless
compensated by the degeneracy of $H_{0}$, since $\left[H_{0},\tau^{z}\sigma^{z}\right]=0$.
Defining the Hamiltonians (using $\Delta_{x}=\Delta\cos\varphi$ and
$\Delta_{y}=\Delta\sin\varphi$ with $\varphi$ the superconducting
phase)
\begin{equation}
H_{p\sigma}=\left(\begin{array}{cc}
\sigma v_{F}(p-\sigma p_{F}) & \sigma\Delta e^{i\varphi}\\
\sigma\Delta e^{-i\varphi} & -\sigma v_{F}(p-\sigma p_{F})
\end{array}\right)
\end{equation}
with $\sigma=\pm1$ representing the two sectors (\emph{i.e.}, the
eigenvalues) of $\tau^{z}\sigma^{z}$. One mixes these sectors when
showing that
\begin{multline}
\Phi_{\pm}=\alpha_{1}\left(\begin{array}{c}
\pm e^{i\varphi}\\
e^{i\gamma}
\end{array}\right)e^{ip_{F}x/\hbar}e^{\pm\frac{x}{\xi}\sin\gamma}\\
+\alpha_{2}\left(\begin{array}{c}
\pm e^{i\varphi}\\
e^{-i\gamma}
\end{array}\right)e^{ip_{F}x/\hbar}e^{\mp\frac{x}{\xi}\sin\gamma}+\\
+\alpha_{3}\left(\begin{array}{c}
\mp e^{i\varphi}\\
e^{i\gamma}
\end{array}\right)e^{-ip_{F}x/\hbar}e^{\mp\frac{x}{\xi}\sin\gamma}+\\
\alpha_{4}\left(\begin{array}{c}
\mp e^{i\varphi}\\
e^{-i\gamma}
\end{array}\right)e^{-ip_{F}x/\hbar}e^{\pm\frac{x}{\xi}\sin\gamma}\label{eq:exact_state_Phi}
\end{multline}
is a superposition of the solutions of $H_{p+}$ and $H_{p-}$ at
the same energy. The index notation in $\Phi_{\pm}$ represents the
two different energies $\pm\Delta\cos\gamma$, and the $\alpha_{i}$
are constants. We parametrized $\varepsilon/\Delta=\cos\gamma$ for
the energies below the gap. The solutions above the gap are found
by the substitution $\gamma\mapsto-i\eta$ such that
the real exponential become some plane-wave with wave-vector parametrized
as $q\xi=\sinh\eta$ and $\varepsilon/\Delta=\cosh\eta$, and thus
$(\varepsilon/\Delta)^{2}-(q\xi)^{2}=1$ corresponds
to a relativistic dispersion relation above the gap. Below the gap
the dispersion relation parametrizes a circle.

\subsection{Midgap states for a semi-infinite wire}

We calculate the midgap state at the interface of a semi-infinite wire in the
half-line $x>0$ with a vacuum located at $x<0$ for $\varphi=0$.  We
essentially follow Ref.~\onlinecite{Sengupta2001}.  This simple example of the
Andreev scattering formalism allows us to explicitly construct the second
quantized version of the Majorana mode, as is usually discussed in literature.
This might be useful for some readers, since the pure wave-function formalism
is not so widely used when discussing Majorana mode physics.

We can concentrate on the positive energy eigenstates $\Phi_{+}$
only from Eq.~(\ref{eq:exact_state_Phi}). Only the exponential decaying
waves must be considered in the semi-infinite geometry. At the $x=0$
interface, the wave function going to the left must be equal to the
right moving wave, since there is a particle vacuum in the $x<0$
space. Then, we impose $\Phi(x=0)=0$. This leads to the
wave-function
\begin{equation}
\Phi_{0}=2\alpha i\left(\begin{array}{c}
1\\
-i
\end{array}\right)e^{-x/\xi}\sin (k_{F}x)
\end{equation}
where the amplitude of the normalization constant $\alpha=Ne^{i\phi}$
is determined by the normalization condition as $N=(2\xi)^{-1/2}$
(we separate the scales $L\gg\xi\gg2\pi k_{F}^{-1}$), whereas the
phase convention is given by the necessity for the spinor $\Phi_{0}$
to describe a real (self-adjoint) solution of the particle-field operator
(second quantized version of the Bogoliubov-de Gennes formalism)
for symmetry reason, in particular since $\left\{ H_{0},\mathcal{P}\right\} =0$.
Then, we choose $\phi=-\pi/4$. This leads to
\begin{equation}
\Phi_{0}=\sqrt{\dfrac{2}{\xi}}\left(\begin{array}{c}
e^{i\pi/4}\\
e^{-i\pi/4}
\end{array}\right)e^{-x/\xi}\sin (k_{F}x)\label{eq:Phi_0_p_wire}
\end{equation}
such that the second-quantized operator representation is (the second
quantized version of the spinor are represented by hats, and $c$
and $c{}^{\dagger}$ are the usual annihilation and creation operator
for fermionic particles)
\begin{multline}
\hat{\Phi}_{0}(x)=\dfrac{1}{\sqrt{\xi}}\int dx\Biggl\{
e^{-x/\xi}\sin (k_{F}x)\\
\times 
\left[e^{i\pi/4}c(x)+e^{-i\pi/4}c{}^{\dagger}(x)\right]\Biggr\}
\end{multline}
and we clearly have $\hat{\Phi}{}_{0}^{\dagger}=\hat{\Phi}_{0}$.
We also remark that $\Phi_{0}=\mathcal{K}\tau^{x}\Phi_{0}$, and is
thus invariant under the particle-hole symmetry of the model. Finally,
note that the mode we have found is a zero energy mode $\varepsilon=\Delta\cos\gamma=0$. 

As a final remark for this section, note that the presence of the Fermi scale
$k_{F}x$ is mandatory for the function $\Phi(x)$ to be an explicit
wave-function satisfying the proper boundary condition $\Phi(0)=0$.  When
$\sin(k_{F}x)$ is neglected in the above expressions, it is not possible to
attribute a momentum to the wave-function.  Said differently, omitting the
$\sin(k_{F}x)$ factor leads to unphysical imaginary eigenvalues of the
momentum operator (which is not Hermitian for wavefunctions with $\Phi(0) \neq
0$).  Here, it is easy to show that the wave-function $\Phi(x)$ minimizes the
Heisenberg uncertainty relation.

\subsection{Wire of finite length $L$\label{sub:Finite-wire}}

We now discuss the situation of a finite length superconducting wire
in the region $0\leq x\leq L$ surrounded by a vacuum. We will calculate
the midgap states in addition to the excited states at energies $\varepsilon\geq\Delta$
above the gap. Then, we simplify the problem in the case of a long
wire $\ell=L/\xi\gg1$, when one can focus on only half of the Majorana
states and when the excited states reduce to sine like wave-functions.

\begin{figure}
\centering
\includegraphics[width=0.95\linewidth]{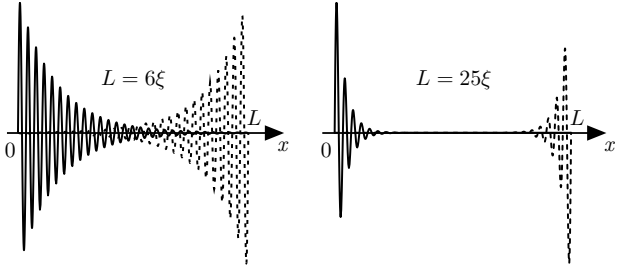}
\caption{Plot of the wave-functions $u_{+}(x/L)\pm u_{-}(x/L)$ given in
Eq.~(\ref{eq:Majorana_modes_LR}) for different length of the $p$-wave
superconductivity wire: the left figure is for $\ell=L/\xi=6$ whereas the
right figure corresponds to $\ell=25$.  The two modes, located on the left and
right edge of the wire, are clearly separated on the second picture.  The
wave-functions $v_{+}\pm v_{-}$ look the same, and are therefore not
represented.  We choose $k_{F}L=180$ for both plots.
\label{fig:Majorana_wave_function}}
\end{figure}

Since we discuss a finite wire geometry, the full solution
(\ref{eq:exact_state_Phi}) must be used.  The geometry imposes
$\Phi(x=0)=\Phi(x=L)=0$.  One obtains
\begin{equation}
(\Xi-1)(\Xi+1)=0\;\;;\;\;\Xi=\dfrac{\sinh(\ell\sin\gamma)\cos\gamma}{\sin(k_{F}L)\sin\gamma}
\end{equation}
for the dispersion relation.  For a given wire length $\ell$ and a given Fermi
momentum $k_{F}$, the dispersion relation gives two modes $\gamma_{\pm}$
corresponding to $\Xi=\pm1$, respectively.  The associated $\alpha_{i}$ are
\begin{equation}
\begin{cases}
\alpha_{1} & =-\left(\Xi e^{-ik_{F}L}+e^{-\ell\sin\gamma}\right)\\
\alpha_{2} & =\Xi e^{-ik_{F}L}+e^{\ell\sin\gamma}\\
\alpha_{3} & =\Xi e^{ik_{F}L}+e^{\ell\sin\gamma}\\
\alpha_{4} & =-\left(\Xi e^{ik_{F}L}+e^{-\ell\sin\gamma}\right)
\end{cases}\label{eq:exact_solutions_amplitude}
\end{equation}
We obtain then 
\begin{equation}
\Phi=\dfrac{1}{N}\sum_{\pm}\left(\begin{array}{c}
u_{\pm}\\
v_{\pm}
\end{array}\right)\label{eq:exact_Phi}
\end{equation}
for the eigenmodes with 
\begin{multline}
\dfrac{u_{\pm}}{i}=\pm\sinh\left(\dfrac{x-L}{\xi}\sin\gamma_{\pm}\right)
\sin(k_{F}x)\\
-\sinh\left(\dfrac{x}{\xi}\sin\gamma_{\pm}\right)\sin\left(k_{F}(x-L)\right)
\end{multline}
and 
\begin{multline}
v_{\pm}=\cosh\left(\dfrac{x-L}{\xi}\sin\gamma_{\pm}\right)\sin(k_{F}x)\sin\gamma_{\pm}\\
\pm\cosh\left(\dfrac{x}{\xi}\sin\gamma_{\pm}\right)\sin\left(k_{F}(x-L)\right)\sin\gamma_{\pm}\\
-\sinh\left(\dfrac{x-L}{\xi}\sin\gamma_{\pm}\right)\cos(k_{F}x)\cos\gamma_{\pm}\\
\mp\sinh\left(\dfrac{x}{\xi}\sin\gamma\pm\right)\cos\left(k_{F}(x-L)\right)\cos\gamma_{\pm}
\end{multline}
and the total wave function is a superposition of the two spinor with
indices $\pm$. The functions $u_{+}\pm u_{-}$ are represented on
Fig.~\ref{fig:Majorana_wave_function}. 
 The other functions are similar, and are thus not represented.

For a long wire, the dispersion relation gives $\gamma_{\pm}\sim\pi/2\pm e^{-\ell}$,
corresponding to a zero-energy mode up to the exponential correction
describing a pair of solutions. This leads to two spinors 
\begin{equation}
\Phi_{L,R}=\dfrac{1}{N_{L,R}}\left(\begin{array}{c}
u_{+}\pm u_{-}\\
v_{+}\pm v_{-}
\end{array}\right)\label{eq:Majorana_modes_LR}
\end{equation}
localized on the left and on the right of the wire, respectively.
Adjusting the norm and the phase of the spinor exponentially decaying
to the right, $\Phi_{L}\approx\Phi_{0}$ as found in Eq.~(\ref{eq:Phi_0_p_wire}).

We now discuss the excited states in the real space representation.
They are given by the substitution $\gamma\mapsto-i\eta$ in all
the previous expressions. It consists essentially in changing all
hyperbolic functions to trigonometric ones for the functions with
$\gamma$ argument; the functions with $k_{F}L$ are obviously not
changed. The dispersion relation reads $(X-1)(X+1)=0$
with 
\begin{equation}
X=\Xi(\gamma=-i\eta)=\dfrac{\sin(\ell\sinh\eta)\cosh\eta}{\sin(k_{F}L)\sinh\eta}\label{eq:excited_states_dispersion}
\end{equation}
for instance; and the amplitudes follow from Eq.~(\ref{eq:exact_solutions_amplitude})
replacing $\Xi$ by $X$.

\begin{figure}
\centering
\includegraphics[width=0.95\linewidth]{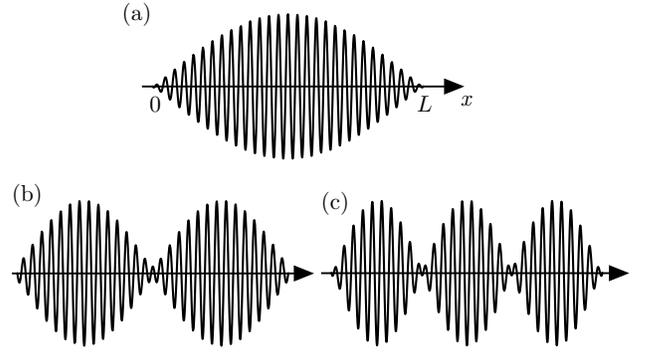}
\caption{Plot of the wave-functions $u_{+}+u_{-}$ for the first excited modes
$n=1$ (a) $n=2$ (b), and $n=3$ (c) computed from Eq.~(\ref{eq:excited_modes})
and Eq.~(\ref{eq:exact_state_Phi}) with the coefficients
(\ref{eq:exact_solutions_amplitude}) for a wire of length $\ell=L/\xi=55$.
The limit as the sine function is clearly demonstrated.  The wave-functions
$u_{+}-u_{-}$ and $i(v_{+}\pm v_{-})$ look the same, and are therefore not
represented.  We choose $k_{F}L=180$ for all these plots.
\label{fig:excited_states} }
\end{figure}

The function $X(\eta)$ is a cardinal sine for short $\eta$ (when
$\cosh\eta\approx1$) whereas it has accelerating oscillations at large $\eta$
as $\sin(\ell\sinh\eta)$ (when
$\cosh\eta/\sinh\eta=(\tanh\eta)^{-1}\approx1$).  So we can approximate the
first solutions for long wire $\ell\gg1$ as
\begin{equation}
\sinh\eta_{\pm}\approx\dfrac{n\pi}{\ell}\pm(-1)^{n}\sin(k_{F}L)\dfrac{n\pi}{\ell^{2}}\label{eq:excited_modes}
\end{equation}
with $n=1,2,\ldots$, The precision increases with a power law $\ell^{-1}$
only, but it is still sufficient. The term $\ell^{-2}$ shows how
the solutions come in pairs. We can thus combine $u_{+}\pm u_{-}$
and $i(v_{+}\pm v_{-})$ (the factor $i$
is added in front of the $v$'s such that the corresponding wave-function
is real, it corresponds to a global phase factor), where $u_{\pm}$
and $v_{\pm}$ are the up ($u$) and down ($v$) components of the
spinor corresponding to the solutions $\eta_{\pm}$, respectively.
The first excited states are plotted in Fig.~\ref{fig:excited_states}.

In the long wire limit, a good approximation for the excited mode
is just
\begin{equation}
\Phi_{n}(x)=N\left(\begin{array}{c}
-1\\
1
\end{array}\right)\sin (k_{F}x)\sin\left(\dfrac{\pi x}{L}
\right)+\mathcal{O}(\ell^{-1})
\end{equation}
with $N=\sqrt{2/L}$ the norm of the spinor.  This is the one used in the main
text, see $(\ref{eq:excited_states})$.  Note that in the main text, we replace
$\pi x/L\approx n\pi x/L=q$ such that the previous pure sine functions are
valid only for long wires and for energies close to the gap.  We numerically
checked the difference between the approximate solution
(\ref{eq:excited_states}) and the exact ones
(\ref{eq:exact_solutions_amplitude}) (with proper replacement
$\Xi(\gamma)\mapsto X(\eta)$ of course) in term of the interaction matrix
element (\ref{eq:interaction_element}) without finding discrepancy in the long
wire limit $\ell\rightarrow\infty$.

More explicitly, one can show that the complete solutions $\Phi(x)$
for the excited states from Eq.~(\ref{eq:exact_Phi}) (after replacement
of $\gamma\mapsto-i\eta$ of course) and the solution
$\Phi_{0}(x)$ from Eq.~(\ref{eq:Phi_0_p_wire}) satisfies
$\int_{0}^{L}dx\left[\Phi_{0}^{\dagger}(x)\Phi(x)\right]\approx e^{-\ell}$
whereas $\int_{0}^{L}dx\left[\Phi_{0}^{\dagger}(x)\tau^{z}\Phi(x)\right]\approx\ell^{-1/2}$
in the long wire limit. When calculating the overlap of the zero-energy
mode and the excited ones, the exponential decay comes from the neglect
in the expression of $\Phi_{0}(x)$ of the zero-energy
mode situated at the right-end edge of the wire, this latter scaling
as $e^{-\ell}$. The calculation can be done straightforwardly in
the scale separation limit when $\sin^{2}(k_{F}x)\approx1/2$, but this
calculation has no specific interest to be written here, since the
manipulation of the expression (\ref{eq:exact_Phi}) is rather cumbersome.
It nevertheless justifies the use of the interaction element (\ref{eq:interaction_element})
in the main text, in addition to the use of the approximate excited
states (\ref{eq:excited_states}).

\section{Evaluation of Eq.~(\ref{eq:g_clean})\label{sec:Integral-evaluation}}

In this section, we give some details about the evaluation of Eq.~(\ref{eq:g_clean}).
Especially, we comment the absence of specific propagating mode at
a velocity well below the Fermi velocity.

We start by rewriting Eq.~(\ref{eq:g_clean}) in the form\cite{note3}
\begin{multline}
g(\omega,x,t)=\dfrac{2}{\pi\sqrt{\xi}}\int_{-t/2}^{t/2}d\tau\int_{0}^{\infty}d\eta\\
\times\left[e^{-i(\omega-\omega_{\Delta}\cosh\eta)\tau}\tanh\eta\sin\left(\dfrac{x}{\xi}\sinh\eta\right)\right].\label{eq:g_clean_appendix}
\end{multline}

Then we follow asymptotic methods evaluation of integrals.\cite{Bender1978}
We first evaluate the integral over $\eta$, defining
\begin{multline}
I(x<v_{F}\tau)=\int_{0}^{\infty}d\eta\tanh\eta\sin(\tilde{x}\sinh\eta)e^{i\tilde{\tau}\cosh\eta}\\
=\dfrac{1}{2i}\int_{-\infty}^{\infty}d\eta\tanh\eta e^{i\left(\tilde{\tau}\sqrt{1-\tilde{v}^{2}}\cosh\left(\arctanh\tilde{v}+\eta\right)\right)}
\end{multline}
with $\tilde{x}=x/\xi$, $\tilde{\tau}=\omega_{\Delta}\tau$ and
 $\tilde{v}=\tilde{x}/\tilde{\tau}=x/v_{F}\tau$.  The above
expression is valid for $\tilde{v}<1$.  One needs to use
\begin{multline}
I(x>v_{F}\tau)=\\
\dfrac{1}{2i}\int_{-\infty}^{\infty}d\eta\tanh\eta e^{i\left(\tilde{\tau}\sqrt{\tilde{v}^{2}-1}\sinh\left(\arctanh\tilde{v}^{-1}+\eta\right)\right)}
\end{multline}
when $\tilde{v}>1$. These two limits are incompatible in the sense that
$I\sim-\ln\sqrt{\tilde{x}^{2}-\tilde{\tau}^{2}}$ when $\tilde{v}\rightarrow1$. So we have
a first indication that (one of) the dominant contribution for the
complete integral appears in the limit of Fermi velocity propagation
$x\approx v_{F}t$.

For $v_{F}\tau\gg x$, one can deform the integral contour to $\eta=z-\arctanh\tilde{v}+i\arcsin(\tanh z)$
with $z$ the new integration variable. This path goes through the
saddle-point at $\eta_{0}=i\sqrt{\tilde{\tau}^{2}-\tilde{x}^{2}}$. There is
obviously no other complication in the $I(x,\tau)$ integral.
Conventional evaluation then leads to 
\begin{equation}
I(x\ll v_{F}\tau)\approx\sqrt{\dfrac{\pi}{2}}e^{3i\pi/4}e^{i\sqrt{\tilde{\tau}^{2}-\tilde{x}^{2}}}\dfrac{\tilde{x}}{\tilde{\tau}^{3/2}}
\end{equation}
for this integral limit.

The second limit $v_{F}\tau\ll x$ has a stationary point at $\eta_{0}=-\arctanh\tilde{v}^{-1}+i\pi/2$
and its asymptotic
\begin{equation}
I(x\gg v_{F}\tau)\approx-\dfrac{\pi}{\sqrt{2}}e^{-\tilde{x}}+i\sqrt{\dfrac{\pi}{2}}e^{-\tilde{x}}\left(\dfrac{\tilde{\tau}}{\sqrt{\tilde{x}}}+\dfrac{\tilde{\tau}^{2}}{\tilde{x}}\right)
\end{equation}
is easily obtained.  For the moment we obtained a propagating wave-like
behavior at velocity $v_{F}$ for large time $\tau$ and a Majorana localized
wave-packet at position larger than $v_{F}\tau$.  In other words, if an
observer sits at the position $x$, the probability amplitude to find an extra
quasi-particle is exponentially weak for times $\tau<x/v_{F}$ and has a power
law decay on time for longer times.

To calculate the time integral, one uses that $\int_{-t/2}^{t/2}e^{i\omega\tau}d\tau=2\Re\left\{ \int_{0}^{t/2}e^{i\omega\tau}d\tau\right\} $
such that one can convert $J(x,t)=\int_{-t/2}^{t/2}d\tau\left[e^{-i\omega\tau}I(x,\tau)\right]$
into an integral over positive $\tau$ only, since this is the only
regime we calculated before. Note that $J=\pi\sqrt{\xi}g/2$ is just
proportional to the $g(\omega,x,t)$ integral for which
the above trick applies. The integral $J(x,t)=2\Re\left\{ j(x,t)\right\} $
must be split in two parts $j(x,t)=j_{1}(x,t)+j_{2}(x,t)$
with
\begin{align}
j_{1}(x,t) & =\int_{0}^{x/v_{F}}I(x\gg v_{F}\tau)e^{-i\omega\tau}d\tau\propto e^{-x/\xi}
\end{align}
which disappears when one integrates $g^{2}$ for long wire, in the
calculation of the $P_\Gamma(t)$. We will no more discuss this
regime, which can be exactly calculated if required, but it is not
relevant in the limit $\ell\gg1$. The second contribution reads 
\begin{equation}
j_{2}(x,t)=\int_{x/v_{F}}^{t/2}I(x\ll v_{F}\tau)e^{-i\omega\tau}d\tau
\end{equation}
for the propagating wave-like integral. This latter integral can be
evaluated by integration by part, since the dominant contributions
arise at the boundaries. It gives 
\begin{multline}
\omega_{\Delta}j_{2}\approx-\sqrt{\dfrac{\pi}{2}}e^{3i\pi/4}\tilde{x}\left[\dfrac{e^{-i\tilde{\omega} \tilde{x}}}{\tilde{x}^{5/2}}-\right.\\
\left.\dfrac{e^{i\sqrt{(\tilde{t}/2)^{2}-\tilde{x}^{2}}}e^{-i\tilde{\omega} \tilde{t}/2}}{(\tilde{t}/2)^{3/2}}\dfrac{i\sqrt{(\tilde{t}/2)^{2}-\tilde{x}^{2}}}{\tilde{\omega}\sqrt{(\tilde{t}/2)^{2}-\tilde{x}^{2}}-\tilde{t}/2}\right]\label{eq:Int_j2}
\end{multline}
with $\tilde{t}=\omega_{\Delta}t$, $\tilde{x}=x/\xi$, and
$\tilde{\omega}=\omega/\omega_{\Delta}$). The expression is
valid when $x/v_{F}t\gg\sqrt{\omega^{2}-\omega_{\Delta}^{2}}/\omega=v_{g}/v_{F}$,
\emph{i.e.}, when the Fermi velocity is larger than the effective group
velocity $v_{g}$ associated to the noise spectrum density at frequency
$\omega$. In the following we neglect the last contribution of $j_{2}$,
since it is time independent. Eq.~(\ref{eq:Int_j2}) leads to Eq.~(\ref{eq:intermediary_times})
of the main text, after taking twice the real part and neglecting
the first line contribution, which is not time dependent.

In the opposite limit of a large effective group velocity, the integral
$j_{2}$ has a saddle-point. To take into account this saddle-point
obliged to consider the regime 
\begin{equation}
\dfrac{(\tilde{\omega}^{2}-1)^{3/2}}{\tilde{t}}\ll\dfrac{\tilde{x}}{\tilde{t}}\ll\dfrac{\sqrt{\tilde{\omega}^{2}-1}}{\tilde{\omega}}\label{eq:vg_condition}
\end{equation}
which in practice imposes the effective group velocity to be close
to its maximum value $v_{g}\approx v_{F}$, since $\omega/\omega_{\Delta}$
is bounded to $1$ in order to make all the results valid, which means
that there is no excitation frequencies below the gap. In that case,
the integral equals
\begin{equation}
j_{2}(v_{g}\approx v_{F})\approx i\dfrac{\pi}{\sqrt{2}\omega_{\Delta}}\dfrac{e^{-i\tilde{x}\sqrt{\tilde{\omega}^{2}-1}}}{\tilde{\omega}^{3/2}\tilde{x}}\label{eq:Int_vg}
\end{equation}
and thus correspond to an effective wave traveling at the effective
group velocity $v_{g}$ only when $v_{g}\approx v_{F}$ in order for
the condition (\ref{eq:vg_condition}) to be verified, so this regime
never dominates in the final integral. 

We carefully checked this point numerically as well. We never found
a situation when the contribution $j_{2}(v_{g})$ is relevant,
except when $v_{g}\approx v_{F}$, in which case the contribution
(\ref{eq:Int_vg}) is well weaker than the dominant contribution (\ref{eq:Int_j2})
and can be safely discarded, as we do in the main text.

One still has to know the long time behavior of the full integral
$g(\omega,x,t)$, when time is the largest parameter of
the integral. This can be done by rewriting
\begin{multline}
g(\omega,x,t)=\dfrac{4}{\pi\omega_{\Delta}\sqrt{\xi}}\int_{1}^{\infty}dz\\
\times\dfrac{\sin\left(\tilde{x}\sqrt{z^{2}-1}\right)}{z}\dfrac{\sin\left(\tilde{t}\left(\tilde{\omega}-z\right)/2\right)}{\tilde{\omega}-z}\label{eq:Int_J}
\end{multline}
after the time integration is performed. When $\tilde{t}\rightarrow\infty$,
the integral is peaked at $z=\tilde{\omega}$, so the first quotient can be
ejected from the integral for the dominant contribution and the lower
boundary can then be replaced by $-\infty$, and the remaining integral
gives $\pi$. The latter argument is equivalent to saying that $\sin(\omega t)/\pi\omega$
behaves like a delta function $\delta(\omega)$ when $t\rightarrow\infty$.
One obtains then
\begin{equation}
\lim_{\tau\rightarrow\infty}g(\omega,x,\tau)=\dfrac{4}{\omega\sqrt{\xi}}\sin\left(\tilde{x}\sqrt{\tilde{\omega}^{2}-1}\right)\label{eq:g_clean_infinite}
\end{equation}
for the leading term. The next correction term is obtained by an expansion
at $z=1-i\epsilon$ for small $\epsilon$. It gives finally
\begin{multline}
g(x,\left(\omega-\omega_{\Delta})t\gg1\right)\approx\dfrac{4}{\omega\sqrt{\xi}}\sin\left(\tilde{x}\sqrt{\tilde{\omega}^{2}-1}\right)\\
-\dfrac{8}{\omega_{\Delta}\sqrt{\pi\xi}}\dfrac{\tilde{x}}{\tilde{t}^{3/2}}\dfrac{\sin\left(\left(\tilde{\omega}-1\right)\dfrac{\tilde{t}}{2}+\dfrac{\pi}{4}\right)}{\tilde{\omega}-1}\label{eq:g_clean_long}
\end{multline}
for large time. Thus the integral goes to a finite value at infinite
time, oscillating in space with a small wave-vector $\sqrt{\omega^{2}-\omega_{\Delta}^{2}}/v_{F}$
when the noise frequency approaches the gap frequency. On top of these
spatial oscillations, there is some wiggling time behavior with long
waves, too. Eq.~(\ref{eq:g_clean_long}) leads to Eq.~(\ref{eq:long-time})
in the main text.

It is pretty difficult to compare our asymptotic expansions at each step of
the calculation, since all the integrals are difficult to integrate even
numerically.  Nevertheless, to compare our asymptotic results with the exact
integral $J(x,t)$, we neglect the $j_{1}$ contribution as it is exponentially
small; \emph{i.e.}, we compare (twice the real part of) Eq.~(\ref{eq:Int_j2})
in the short time limit and Eq.~(\ref{eq:g_clean_long}) valid for long time
with the numerical evaluation of the complete integral Eq.~(\ref{eq:g_clean}).
Some characteristic curves are given in Fig.~\ref{fig:g_omega}.


\begin{thebibliography}{10}
\bibitem{note1} The previous argument is equivalent to saying that
$\sin^{2}(\omega t)/\pi\omega^{2}t$ behaves like $\delta(\omega)$ when
$t\rightarrow\infty$.
\bibitem{note2} The static contribution $\left\langle x|0\right\rangle $
vanishes exponentially in the space-integration over the half-wire and is thus
neglected from the beginning.
\bibitem{note3} We use the completeness relation $1=\left|0\right\rangle
\left\langle 0\right|+\int\left|q\right\rangle \left\langle q\right|Ldq/\pi$,
the definitions $\varepsilon=\Delta\cosh\eta$ and $q\xi=\sinh\eta$ for the
quasi-particle modes; also, $\sin^{2}(k_{F}x)\approx1/2$ as $k_F$ is large.
\end{thebibliography}
\end{document}